\documentclass[aps,twocolumn,10pt,aps,amsmath,amssymb,nofootinbib,superscriptaddress]{revtex4-2}
\usepackage{epsfig} 
\usepackage{amsmath} 
\usepackage{graphicx}
\usepackage{color}
\usepackage{soul,xcolor}
\usepackage[font=scriptsize]{caption}
\usepackage{braket}
\usepackage{bbold}
\usepackage{subfig}
\usepackage{braket}
\usepackage{titlesec}
\usepackage{placeins}
\usepackage{hyperref}
\usepackage{cleveref}
\usepackage{caption}
\usepackage[bottom]{footmisc}
\begin{document}
\title{Violation of LGtI inequalities in the light of NO$\nu$A and T2K anomaly}

\author{Lekhashri Konwar}
\email{konwar.3@iitj.ac.in}
\affiliation{Indian Institute of Technology Jodhpur, Jodhpur 342037, India}

 \author{Juhi Vardani}
\email{vardani.1@iitj.ac.in}
\affiliation{Indian Institute of Technology Jodhpur, Jodhpur 342037, India}

\author{Bhavna Yadav}
\email{yadav.18@iitj.ac.in}
\affiliation{Indian Institute of Technology Jodhpur, Jodhpur 342037, India}

\begin{abstract}
The recent anomaly observed in NO$\nu$A and T2K experiments in standard three-flavor neutrino oscillation could potentially signal physics extending beyond the standard model (SM). For the NSI parameters that can accommodate this anomaly, we explore the violation of Leggett-Garg type inequalities (LGtI) within the context of three-flavor neutrino oscillations. Our analysis focuses on LGtI violations in scenarios involving complex NSI with $\epsilon_{e\mu}$ or $\epsilon_{e\tau}$ coupling in long baseline accelerator experiments for normal and inverted mass ordering.LGtI violation is significantly enhanced in normal ordering (NO) for $\epsilon_{e\tau}$ scenario for T2K, NO$\nu$A, and DUNE experiment set-up. We find that for inverted ordering (IO), in the DUNE experimental set-up above $8.5$ GeV, the LGtI violation can be an indication of $\epsilon_{e\tau}$ new physics scenario.
%
\end{abstract} 
\maketitle
\section{Introduction}
\label{intro}
The occurrence of neutrino oscillations is a consequence of the superposition principle, which dictates that the flavor eigenstate ($\nu_{\alpha}$) must undergo mixing with the mass eigenstate ($\nu_i$). Due to this mixing, the flavor eigenstate ($\nu_{\alpha}$) can be expressed as a linear combination of mass eigenstates ($\nu_i$). As a neutrino of a specific flavor $\nu_{\alpha}$ travels, it undergoes a proportional change in its mass-state composition from its initial state. Consequently, when detected at a certain distance, it manifests as having a different flavor, denoted as $\nu_{\beta}$. This dynamic transformation in the flavor state constitutes the fundamental essence of neutrino oscillation. These oscillations in the flavor states of neutrinos imply that neutrinos possess non-zero masses. Evidence supporting neutrino oscillations comes from various experiments, including solar, atmospheric, reactor, and accelerator experiments. 

These experiments are specifically designed to measure key oscillation parameters: the two mass squared differences ($\Delta m^2_{21}$, $\Delta m^2_{31}$), as well as the three mixing angles $\theta_{21}$, $\theta_{23}$, and $\theta_{31}$, along with the Dirac CP phase $(\delta_{CP})$. The neutrino oscillation probability can be formulated in terms of these parameters. However, it is important to note that the sign of $\Delta m^2_{31}$ and the precise value of the CP phase $(\delta_{CP})$ remain undetermined.

While traversing the earth's crust, neutrinos experience the influence of SM matter potential referred to as the matter effect. In addition, one can also have non-standard interactions characterized by deviation from standard neutrino interactions with matter, and these deviations can be expressed through effective coupling $(\epsilon_{\alpha\beta})$. In the presence of non-standard interactions (NSI), neutrino oscillation probability can be described in terms of NSI parameters. As a result, their effects can be measured in upcoming long-baseline neutrino experiments. The effective coupling can be real or complex. 

In recent two long baseline accelerator experiments, NO$\nu$A and T2K released data for $\delta_{CP}$ \cite{Himmel, Dunne}. 
Their experimental result depicts a mismatch for normal ordering. NO$\nu$A prefers $\delta_{CP}$ $\sim$ $0.8\pi$, while T2K governs CP phase value $1.4\pi$. These two assessments are in disagreement, exceeding $90\%$ C.L. with two degrees of freedom. While results for inverse ordering are matched, the non-zero value of the CP phase parameter in T2K and the mismatch with NOvA in the mass matrix suggest that complexity is present not only in the mass matrix it can be also in new physics.

Several new physics solutions have been proposed to address the anomalies observed in T2K and NO$\nu$A, with a notable contribution from complex NSI \cite{Denton:2020uda,Chatterjee:2020kkm,Cherchiglia:2023ojf}. The resolution of the discrepancy between NOvA and T2K crucially involves a new CP phase, particularly in the context of flavor-changing neutral current NSI, denoted as $\epsilon_{\alpha\beta}$ where $\alpha$ and $\beta$ are not equal. In ref. \cite{Chatterjee:2020kkm}, the off-diagonal NSI parameters $\epsilon_{e \mu}$ and $\epsilon_{e \tau}$ are considered, with a focus on their impact on the observed anomaly. The introduction of these NSI parameters provides an effective means to study the novel CP phase. The parameters $\epsilon_{e \mu}$ and $\epsilon_{e \tau}$ are best explored through long baseline accelerator appearance experiments, while the investigation of $\epsilon_{ \mu \tau}$ is found to be most effective when utilizing atmospheric neutrinos.

{\color{black}The existence of new physics also holds the potential to impact quantum correlations in neutrino oscillations. The quantification of quantum correlations, explored within the framework of quantum information theory, has garnered considerable attention in recent years. This also includes applications in high-energy physics systems, notably neutrino and mesonic systems, where various methodologies and tools associated with quantum information theory have been extensively explored,  see for e.g., }\cite{Blasone:2007wp,Blasone:2007vw,Banerjee:2014vga,Alok:2014gya,Banerjee:2015mha,Formaggio_2016,Fu:2017hky,Naikoo:2017fos,Naikoo:2018vug,Dixit:2018gjc,Naikoo:2019eec,Dixit:2019swl,Shafaq:2020sqo,Sarkar:2020vob,Ming:2020nyc,Blasone:2021mbc,Yadav:2022grk,Blasone:2022iwf,Chattopadhyay:2023xwr,Blasone:2023gau,Caban:2007je,Alok:2024amd,Capolupo:2018hrp,IceCube:2023gzt}.  The influence of new physics can manifest through both temporal and spatial quantum correlations within a system. Bell's inequality primarily deals with corrections in measurements conducted on spatially distant systems \cite{bell}. In a parallel manner, Leggett and Garg developed a test analogous to Bell's inequality, but with a focus on corrections in measurements made on a system at different points in time \cite{Leggett1985}.
The Leggett-Garg theorem is based on the assumptions of macrorealism (MR) and noninvasive measurement (NIM). A violation of the Leggett-Garg inequality implies that either the system cannot be realistically described, or it is impossible to measure the system without disturbing it. However, systems driven by principles of quantum mechanics unequivocally diverge from these ~\cite{Leggett1985,Emary_2013}.  A less stringent criterion, referred to as the stationarity condition, is employed to formulate Leggett-Garg-type inequalities (LGtI) as a substitute for the NIM condition \cite{Huelga1995}. Since neutrino measurements undermine the NIM assumption, such inequalities are preferred to analyze temporal correlation in neutrinos \cite{Naikoo:2017fos,Naikoo:2019eec}. Violation of temporal correlation in neutrino oscillations is verified using the oscillation MINOS and Daya-Bay experiments \cite{Formaggio_2016,Fu:2017hky}. In this study, we explore the impact of NSI on the potential violation of LGtI. To do so, we specifically examine complex NSI parameters, namely $\epsilon_{e\mu}$ and $\epsilon_{e\tau}$, which have been demonstrated to accommodate the existing anomalies observed in NO$\nu$A and T2K experiments \cite{Chatterjee:2020kkm}. 

The structure of this work is outlined as follows. In the next section, we describe the formalism, beginning with a discussion on SM matter effects, followed by a concise description of the NSI employed in our study. Subsequently, we provide a brief definition of LGtI pertinent to neutrino oscillations. In the subsequent section, we present and discuss our results. The conclusions drawn from our findings are encapsulated in the final section.
\\
\\
\section{Formalism}
\label{sec:2}
\emph{Neutrino Oscillation in matter:}
The minute mass of neutrinos results in the mixing of mass eigenstates, leading to the phenomenon of neutrino oscillation, where a neutrino transitions from one flavor to another over time. This implies that the flavor eigenstate ($\nu_\alpha$) can be represented as a linear combination of mass eigenstates ($\nu_i$) in the following manner:
\begin{equation}
\ket{\nu_\alpha}  = \sum_{i=1}^{3} U_{\alpha i}\ket{\nu_i}\,.
\end{equation}  
Here $U_{\alpha i}$ represents the elements of the Pontecorvo-Maki-Nakagawa-Sakata (PMNS) matrix. The mixing matrix U can be parameterized by three mixing angles $(\theta_{12},\theta_{23},\theta_{31})$, and a CP violating phase $\delta_{cp}$. The time-evolved mass eigenstate at time $t$ can be expressed as $\ket{\nu_i(t)} = e^{-\iota \mathcal{H}_mt}\ket{\nu_i} = e^{-\iota E_it}\ket{\nu_i}$, where $E_i$ is the eigenvalue corresponding to the mass eigenstate $\ket{\nu_i}$ and $U_m=e^{-\iota \mathcal{H}_mt}$.

The time-evolved flavor eigenstate at time $t$ can then be written as:
\begin{eqnarray}
\ket{\nu_{\alpha}(t)} &=& \sum_{i=1}^{3} U_{\alpha i} e^{-\iota E_it} \ket{\nu_i}\nonumber\\
        && = \sum_{i,\beta=1}^{3} U_{\alpha i} e^{-\iota E_it}U^{*}_{\beta i}\ket{\nu_\beta}  
\end{eqnarray}
\begin{equation}
    =U_f(t)\ket{\nu_\beta}.
\end{equation}
Here $\ket{\nu_i}$ and $\ket{\nu_\beta}$ represent the mass and flavor eigenstates at time $t=0$. $U_f(t)= U e^{-\iota H_mt}U^{-1}$ is the time-evolution operator. The Hamiltonian $\mathcal{H}_m$ of $\nu$ in the mass basis, when the neutrino propagates through matter, is given by:
\begin{equation}
    \mathcal{H}_m =\begin{pmatrix}
        E_1 & 0 & 0\\
        0 & E_2 & 0\\
        0& 0& E_3
    \end{pmatrix} 
    +U^{\dagger}\begin{pmatrix}
        A & 0 & 0\\
        0 & 0 & 0\\
        0 & 0 & 0
    \end{pmatrix}U .
\end{equation}
Here $A=\pm \sqrt{2}G_{f}N_{e}$ represents the standard matter potential, where $G_f$ is the Fermi constant and $N_e$ is the electron number density in matter. The matter potential is positive for neutrinos and negative for antineutrinos. In the ultrarelativistic limit when $t=L$, the evolution operator in the mass basis is expressed as:
\begin{equation}
\begin{split}
U_m(L)&= e^{-\iota\mathcal{H}_mL}\\& = \phi \sum_{a=1}^3 e^{-\iota L\lambda_a}\frac{1}{3\lambda^2_a+c_1}[(\lambda_a^2+c_1)I+\lambda_a T+T^2].
\end{split}
\end{equation}
Similarly, the flavor evolution can be written as 
\begin{eqnarray}
   U_f(L)= e^{-\iota\mathcal{H}_fL}= U e^{-\iota\mathcal{H}_mL} U^{-1}~~~~~~~~~~~~~~~~~~~~~~~~~~~~\\=\phi \sum_{a=1}^3 e^{-\iota L\lambda_a}\frac{1}{3\lambda^2_a+c_1}[(\lambda_a^2+c_1)I+\lambda_a\Tilde{T} +\Tilde{T^2}]\nonumber,
 \end{eqnarray} 
where $\Tilde{T}=UTU^{-1}$, $T= \mathcal{H}_m- (tr\mathcal{H}_m)\frac{I}{3}$, $\lambda_a(a=1,2,3)$ are eigen values of T matrix, and $c_1= det(T)Tr(T^{-1)}$.

\emph{Non-Standard Interaction (NSI) in Neutrino Oscillation:} 
In addition to standard interactions, neutrino oscillations experience NSI with matter. NSI can be either charged or neutral. Charged-current (CC) NSI involves neutrinos interacting with matter fields (electrons, up quarks, and down quarks), affecting the production and detection of neutrinos. On the other hand, neutral-current (NC) NSI influences the propagation of neutrinos in matter. As a result, both types of interactions can manifest in different ways across various neutrino experiments. This can be represented by the 6-dimensional fermion operator Lagrangian as follows: 
\begin{eqnarray}
    \mathcal{L}_{CC-NSI}=2\sqrt{2}G_F\epsilon_{\alpha \beta}^{ff^{\prime},L}(\bar{\nu_\alpha}\gamma^\mu P_Ll_\beta)(\bar{f'}\gamma_\mu P_Lf)\nonumber \\+ 2\sqrt{2}G_F\epsilon_{\alpha \beta}^{ff',R}(\bar{\nu_\alpha}\gamma^\mu P_Ll_\beta)(\bar{f'}\gamma_\mu P_Rf) ,
 \end{eqnarray} 
\begin{eqnarray}
    \mathcal{L}_{NC-NSI}=2\sqrt{2}G_F\epsilon_{\alpha \beta}^{f,L}(\bar{\nu_\alpha}\gamma^\mu P_L\nu_\beta)(\bar{f}\gamma_\mu P_Lf)
\nonumber\\+ 2\sqrt{2}G_F\epsilon_{\alpha \beta}^{f,R}(\bar{\nu_\alpha}\gamma^\mu P_L\nu_\beta)(\bar{f}\gamma_\mu P_Rf).
 \end{eqnarray}
Here $P_{L/R}=\frac{1\mp \gamma_5}{2}$ are left and right-handed chirality operators. The dimensionless coefficient $\epsilon_{\alpha\beta}$ quantifies the strength of the NSI in comparison to the weak interaction coupling constant $G_{f}$, i.e., $\epsilon_{\alpha, \beta}^{ff', L/R}$  $\approx(G_x/G_f$), while $\alpha$ and $\beta$ correspond to various lepton flavors $\{\alpha,\beta\} \in \{e,\mu, \tau\}$. For NC interaction, $f$=e,u,d fermions, whereas for CC interaction, $f,f^\prime$=u,d quarks. In our analysis, the effect of incoherent scattering is neglected, and the density of the earth is assumed to be 2.84g/cc for NO$\nu$A and DUNE with $A=1.08\times 10^{-13}$eV, whereas 2.6g/cc for T2K with $A=0.98\times 10^{-13}$eV \cite{Denton:2020uda}. In the presence of NSI, the interaction Hamiltonian on a mass basis is modified as follows:
\begin{equation}
    \mathcal{H}_{tot} =\mathcal{H}_{vac}+\mathcal{H}_{mat}+ \mathcal{H}_{NSI},
\end{equation}
where ${H}_{vac}$ and ${H}_{mat}$ is the Hamiltonian for vacuum and the matter respectively, and the NSI contribution is shown by ${H}_{NSI}$. The combined effect of all scenarios is shown by the total Hamiltonian, ${H}_{tot}$, which can be represented as
\begin{equation}
  \begin{pmatrix}
        E_1 & 0 & 0\\
        0 & E_2 & 0\\
        0& 0& E_3
    \end{pmatrix} 
    +U^{\dagger}A\begin{pmatrix}
        1+\epsilon_{ee}(x) & \epsilon_{e\mu}(x)& \epsilon_{e\tau}(x)\\
        \epsilon_{\mu e}(x) & \epsilon_{\mu\mu}(x) & \epsilon_{\mu\tau}(x)\\
        \epsilon_{\tau e}(x) & \epsilon_{\tau\mu}(x) & \epsilon_{\tau\tau}(x)
    \end{pmatrix}U.
\end{equation}
Here, all $\epsilon_{\alpha\beta}$ parameters are NSI parameters that can be expressed as
\begin{equation}\label{ep}
    \epsilon_{\alpha\beta}= \epsilon_{\alpha\beta}^e+(2+Y_n)\epsilon_{\alpha\beta}^u+(1+2Y_n)\epsilon_{\alpha\beta}^d, 
\end{equation} 
where $Y_n=N_{n}/N_{e}$, which is the ratio of neutron number density $(N_{n})$ to electron number density $(N_{e})$. Eq. \ref{ep} can be obtained after following two conditions: one is the necessity of charge neutrality in the matter, which suggests $N_{p}=N_{e}$, while the other condition is using the quark structure of proton and neutron. From the hermiticity condition, the diagonal terms are real; however, off-diagonal terms can be complex.

\emph{Leggett Garg type inequalities (LGtI):}
Leggett-Garg inequalities (LGI) are a measure of quantum correlation analogous to Bell inequalities that concern spatial quantum correlation measures. However, LGI is a temporal quantum correlation measure that measures the system's correlation at various instance of time \cite{Leggett1985}. LGI is based on two assumptions: $(1)$ Macro-realism (MR) $(2)$ Noninvasive measurement (NIM). According to macro-realism, a macroscopic system that can access two or more macroscopically distinct states will always be available in one of these states. NIM asserts that it is possible to perform a measurement on a system without even disturbing its dynamics. LGI manifests the level up to which quantum mechanics is applied to many-particle systems that exhibit decoherence, i.e., it manifests macroscopic coherence. LGI also tests the notion of realism, implying that a physical system possesses a complete set of definite values for various parameters prior to and independent of measurement. Realism introduces the concept of hidden variable theory. The violation of LGI suggests that such hidden variable theory cannot be used as an alternative tool for the time evolution of quantum mechanics. The LGI parameter $K_{n}$ can be expressed in terms of a linear combination of autocorrelation functions, as \cite{Naikoo:2017fos}:
\begin{equation}
    \sum_{i=1,j=i+1}^{i=n-1}  C\left ( t_{i} ,t_{j}\right ) - C\left ( t_{1},n \right ),
\end{equation}
with $C(t_i,t_j)=\langle\hat{Q}(t_i)\hat{Q}(t_j)\rangle$, where $C(t_i,t_j)$ is  autocorrelation function and the expectation value of an operator can be written as
$\langle\hat{Q}(t_i)\hat{Q}(t_j)\rangle$  $=$ $Tr[\{\hat{Q}(t_i),\hat{Q}(t_j)\}\rho(t_0)$].
Here, the average is taken with respect to $\rho(t_{0})$, initial state of the system at time $t=0$, $\hat{Q}$ is generic dichotomic operator, i.e., $\hat{Q}=\pm 1$, $\hat{Q}^+= Q$ and $\hat{Q}^2=1$ with $\hat{Q}= 1$ if system is in target state; otherwise, $\hat{Q}=- 1$. For $n\geq 3$, if realism is satisfied, then the LGI parameter $K_n\leq n-2$. For n$=$3, LG inequality can be written as \begin{equation}
 K_3= C(t_1, t_2)+C(t_2,t_3)-C(t_1,t_3) \leq 1.
\end{equation}
Since $C(t_2,t_3)$ depends on the complete measurement of the system, it destroys the NIM postulate. To overcome this, the NIM postulate is replaced by a weaker condition of stationarity. After applying the weaker condition of stationarity, the correlation function $C(t_{i},t_{j})$ depends on the time difference. Under the assumption that time interval $t_2-t_1$ = $t_3-t_2$ = t and $t_1=0$ the LGtI parameter $(K_3)$ can be written as \cite{Emary_2013} :
\begin{equation}
    K_3= 2C(0,t)-C(0,2t)\leq 1.
\end{equation}
For three-flavor neutrino oscillation, $K_3$ can be expressed in terms of neutrino oscillation probability as \cite{Formaggio_2016,Naikoo:2019eec}:
\begin{equation}
    K_3= 1+2P_{\alpha\beta}(2t,E)- 4P_{\alpha\beta}(t,E).
\end{equation}
In the ultrarelativistic limit $t=L$ the LGI  parameter $K_3$ is given as:
\begin{equation}
   K_3= 1+2P_{\alpha\beta}(2L,E)- 4P_{\alpha\beta}(L,E).
\end{equation}
Which shows experimental feasibility of LG function on using condition $P_{\alpha, \beta}(2L,E)$ = $P_{\alpha, \beta}(L,\Tilde{E})$ with suitable choice of E and $\Tilde{E}$ \cite{Formaggio_2016}.
\\
\section{Results and Discussion}
\label{sec:3}
In this section, we examine the impact of NSI scenarios involving $\epsilon_{e \mu}$ and $\epsilon_{e \tau}$ on the temporal measurement of quantum correlations within the neutrino system, considering three distinct experimental setups: $\left( 1 \right)$ T2K (E $\approx$ 0 to 6 GeV, L= 295 km), $\left( 2 \right)$ NO$\nu$A (E $\approx$ 1 to 10 GeV, L= 810 km), and $\left( 3 \right)$ DUNE (E $\approx$ 1 to 10 GeV, L= 1300 km).

\begin{table}[htb!] 
    \begin{tabular}{|c | c| c |}
    \hline
    \hline
     Input Parameters & ordering & 1$\sigma$ range\\
    \hline
    $\delta m^2/10^{-5}$ e$v^2$ & NO & [7.20 - 7.51]\\
      & IO & [7.20 - 7.51]\\
    \hline
    $\sin^2{\theta_{12}}$ & NO & [0.292 - 0.319]  \\
     & IO & [0.290 - 0.317]  \\
    \hline
    $\Delta m^2/10^{-3}$ e$v^2$ & NO & [2.453 - 2.514]\\
      & IO & [2.434 - 2.495]\\
      \hline
      $\sin^2{\theta_{13}}$ & NO & [0.0214 - 0.0228]  \\
     & IO & [0.0217 - 0.0230]  \\
     \hline
     $\sin^2{\theta_{23}}$ & NO & [0.498 - 0.0565]  \\
     & IO & [0.0517 - 0.0567]  \\
     \hline
     $\delta/ \pi$ & NO & [1.10 - 1.66]  \\
     & IO & [1.37 - 1.65]  \\
     \hline
    \end{tabular}
    \caption{Best fit values of input parameters \cite{Capozzi:2017ipn}.}
    \label{Inputs}
\end{table}

\begin{table}[t!] 
    \centering
    \begin{tabular}{|c|c|c|c|c|}
        \hline
       mass ordering  & NSI & $\epsilon_{\alpha\beta}$ & $\phi_{\alpha\beta}/ \pi$ & $\delta_{CP}/\pi$ \\
       \hline
        NO & $\epsilon_{e\mu}$ & [0.057 - 0.22]& [0.88 - 1.9] & [1.17 - 1.79]\\
        IO & $\epsilon_{e\mu}$ & [0 - 0.149] & [0 - 360] &[1.27 - 1.74]\\
        \hline
        NO & $\epsilon_{e\tau}$ & [0.078 - 0.492] & [1.25 - 1.91] & [1.2 - 1.79]\\
        IO & $\epsilon_{e\tau}$ & [0 - 0.4] & [1.02 - 1.97] & [1.24 -- 1.82]\\
        \hline
    \end{tabular}
    \caption{ {\color{black} $1\sigma$ range of NSI parameters and $\delta_{CP}$ used in our analysis} \cite{Chatterjee:2020kkm}.}
    \label{del1}
\end{table}

\begin{figure*}[ht!]
\centering
\includegraphics[width=75mm]{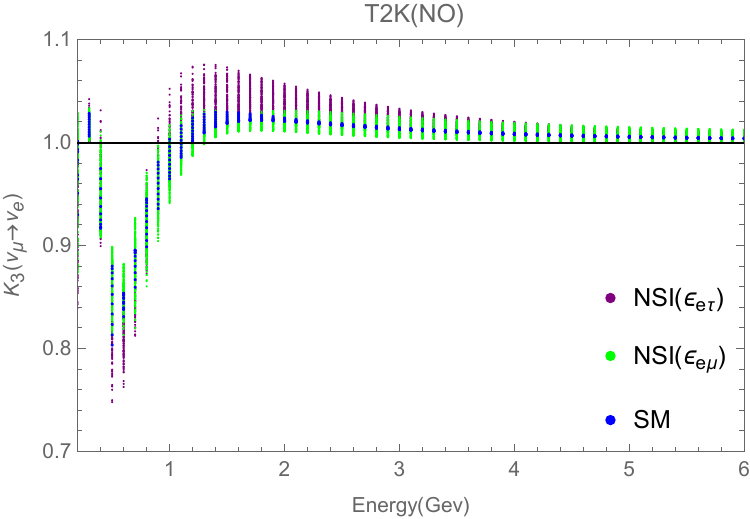}
\includegraphics[width=75mm]{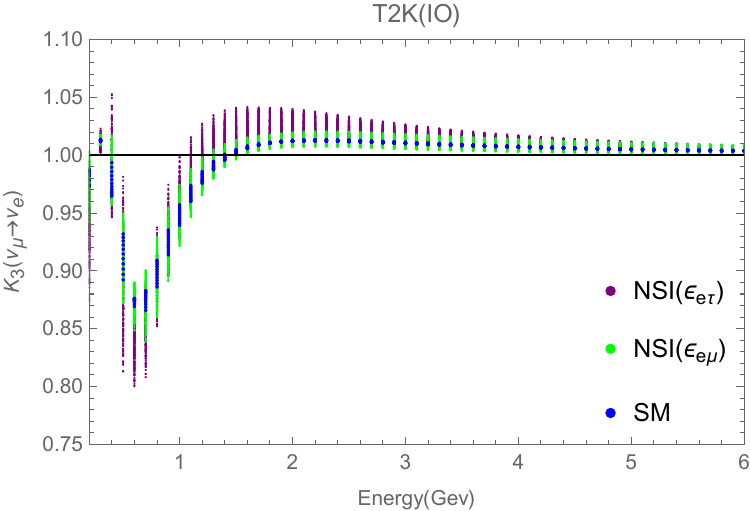}
\\
\includegraphics[width=75mm]{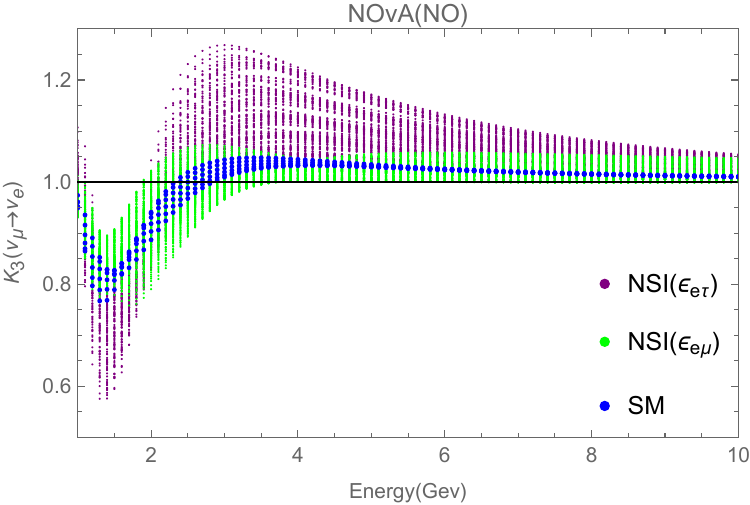}
\includegraphics[width=75mm]{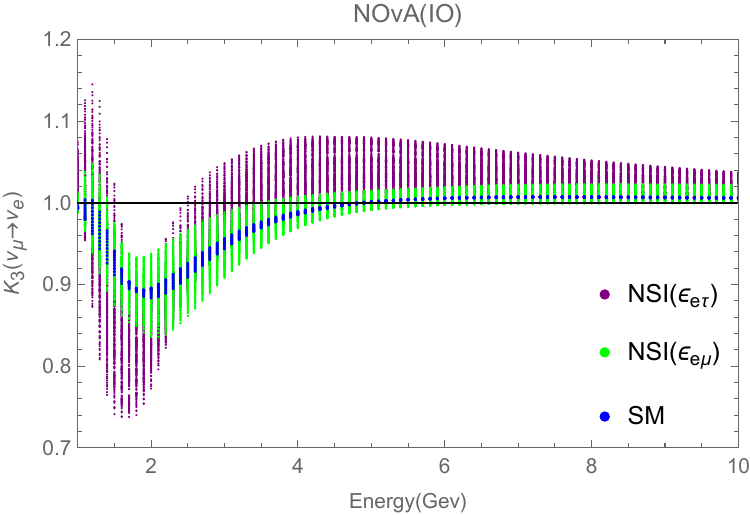}
\\
\includegraphics[width=75mm]{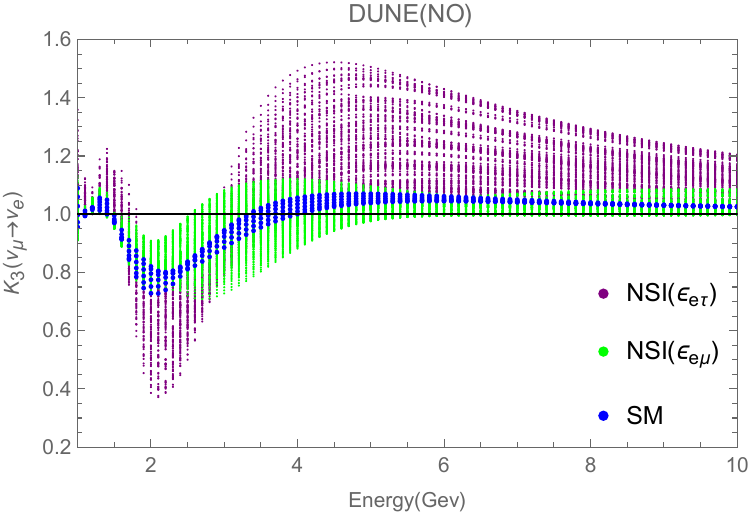}
\includegraphics[width=75mm]{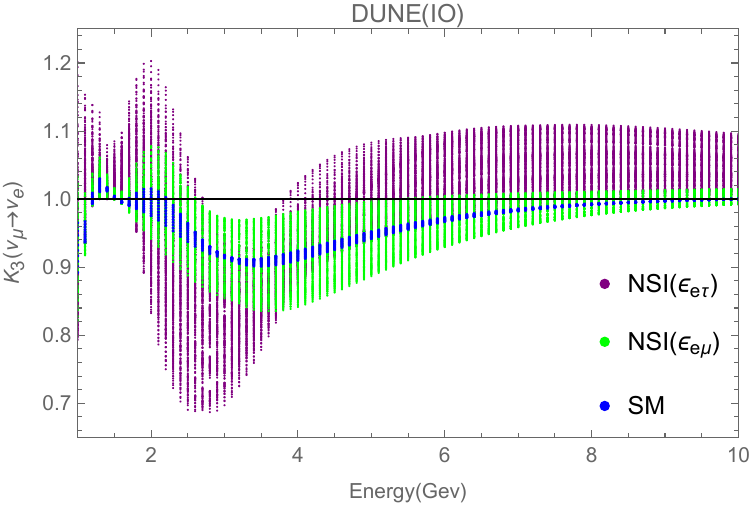}
\caption{ Temporal quantum correlation, quantified in terms of LGtI (denoted as $K_{3}$), is plotted against neutrino energy for T2K, NO$\nu$A, and DUNE experiments.{\color{black} The green band represents LGtI with a $1\sigma$ allowed range of $\epsilon_{e\mu}$, $\phi_{e\mu}$, and cp phase, the purple band represents LGtI with $1\sigma$ allowed range of $\epsilon_{e\tau}$, $\phi_{e\tau}$ and cp phase, and the blue band represents LGtI with $1\sigma$ allowed range of cp phase.} }
\label{fig1}
\end{figure*}

These are long baseline accelerator experiments with neutrino energy of the order of GeV and a baseline length of a few hundred kilometers. These experiments are sensitive to measure mass hierarchy ($\Delta m^2_{31}$), mixing angle ($\theta_{23}$), and CP phase ($\delta_{CP}$).
The value of standard oscillation parameters given in Table \ref{Inputs} has been taken from ~\cite{Capozzi:2017ipn}. The value of NSI parameters has been extracted from global analysis of the appearance and disappearance channels of combined NO$\nu$A and T2K data \cite{Chatterjee:2020kkm}.  {\color{black} The $1\sigma$ range of these parameters are given in Table \ref{del1}. Table \ref{NSI} illustrates the energy range in all three accelerator experiments where $K_{3}$ surpasses its classical bound for three flavour neutrino oscillation for SM and NSI scenarios. }  

 The outcomes for LGtI violation within the $1\sigma$ range in various accelerator experiments are visualized in Fig. \ref{fig1}, with T2K at the top, NO$\nu$A in the middle, and DUNE at the bottom.

{\color{black} In Fig. \ref{fig1}, for the \textbf{T2K} experiment, for the NO, $K_{3}$ is consistently violated for energies above $1.3$ GeV. $K_{3}$ values corresponding to the SM and NSI scenarios exhibit overlapping regions within the error bar. So, the effects are most evident when observing maximum deviations.  In the presence of $\epsilon_{e\tau}$, the violation of $K_{3}$ increases up to $4\%$ as compared to the SM  within the energy range of $1$ GeV to $2$ GeV. For the IO, the violation of $K_{3}$ occurs consistently for an energy range exceeding $1.5$ GeV. However, there is no significant distinction between SM and NSI scenarios.}

\begin{table}[b] 
    \centering
    \begin{tabular}{|c|c|c|c|c |}
        \hline
       Experiment & MO &   Energy in GeV \\
       \hline
        DUNE & NO &[1, 1.5], [3.4, 10], [6, 10]\\
         \hline
       & IO &  [1.2, 1.5], [1.7, 2.2]\\
         \hline
       NO$\nu$A & NO & [3.5, 10]\\
         \hline
         & IO &  [6.5, 10]\\
         \hline
         T2K & NO &  [0.2, 0.4], [1.3, 6] \\
         \hline
         & IO &  [1.5, 6]\\
         \hline
    \end{tabular}
    \caption{{\color{black} Specific range of energy where temporal correlation parameter exceeds its classical limit, \textit{i.e.} $K_3$ is violated for different mass ordering in SM and NSI scenarios.}}
    \label{NSI}
\end{table}

{\color{black} For \textbf{NO$\nu$A} experiment, in the realm of NO, $K_3$ is always violated above $3.5$ GeV. The $K_{3}$ values for SM and NSI scenarios exhibit overlap within the one-sigma confidence region, so the effect is most apparent when observing maximum deviations.  In the presence of $\epsilon_{e\tau}$, the violation of $K_{3}$ increases up to $16\%$ as compared to the SM  within the energy range of $3.8$ GeV to $4.5$ GeV. While the maximum violation can be reached up to $22\%$ for this scenario. For IO, above $6.5$ GeV, $K_{3}$ always shows violation. For energy region $(2.5-3.5)$ GeV only $\epsilon_{e \tau}$ shows violation.}

{\color{black} For the \textbf{DUNE} experiment, in NO, $K_{3}$ is consistently violated for energies above 6 GeV. $K_{3}$ values corresponding to the SM and NSI scenarios exhibit overlapping regions within the one-sigma range. Consequently, discerning the effects is most evident when observing maximum deviations.  In the presence of $\epsilon_{e\tau}$, the violation of $K_{3}$ increases up to $32\%$ as compared to the SM  within the energy range of $3.5$ GeV to $6.5$ GeV. While the maximum violation can be reached up to $40\%$ for this scenario. Whereas violation enhanced up to $5\%$ in the $\epsilon_{e\mu}$ scenario within the energy range of $3.5$ GeV to $4.5$ GeV. For IO, violation of $K_3$ occurs for SM and both NSI scenarios for $(1.2-1.5)$ GeV energy range. Within the error bar, the maximum deviation in violation of $K_3$ for $\epsilon_{e\mu}$ and $\epsilon_{e\tau}$  as compared to SM for $(1.6-2.2)$ GeV energy range is $5\%$ and $15\%$, respectively. For IO above $8.5$ GeV, $K_3$ always shows violation for the $\epsilon_{e\tau}$ scenario, within $1\sigma$ range, while SM does not show violation. Therefore, the measurement of $K_{3}$ has the potential to discriminate among SM and $\epsilon_{e\tau}$ scenario, suggesting a potential indication of new physics.  Additionally, within this range, the maximum enhancement in $K_3$ is $10\%$ for the $\epsilon_{e\tau}$ scenario.}

\begin{figure*}[ht!]
\centering

\includegraphics[width=75mm]{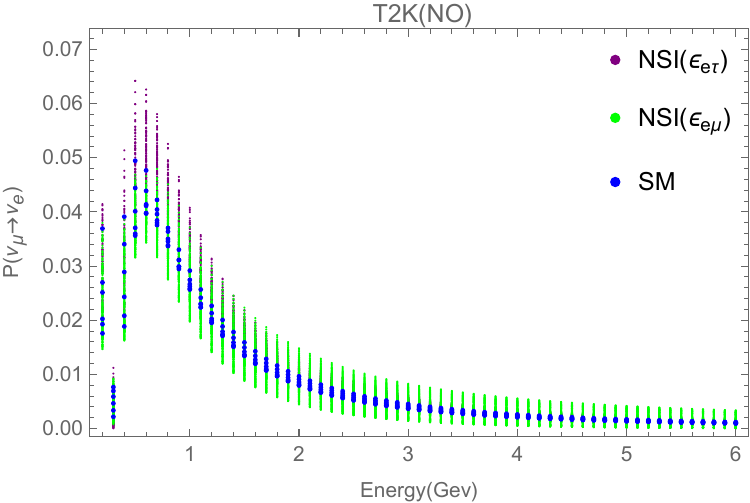}
\includegraphics[width=75mm]{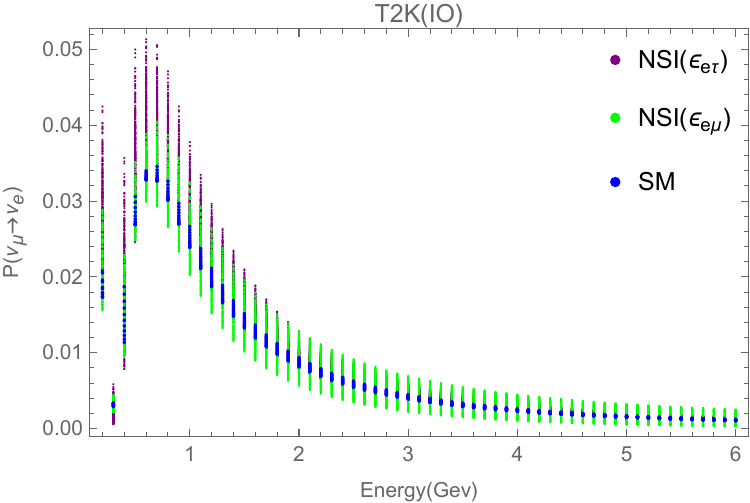}
\\
\includegraphics[width=75mm]{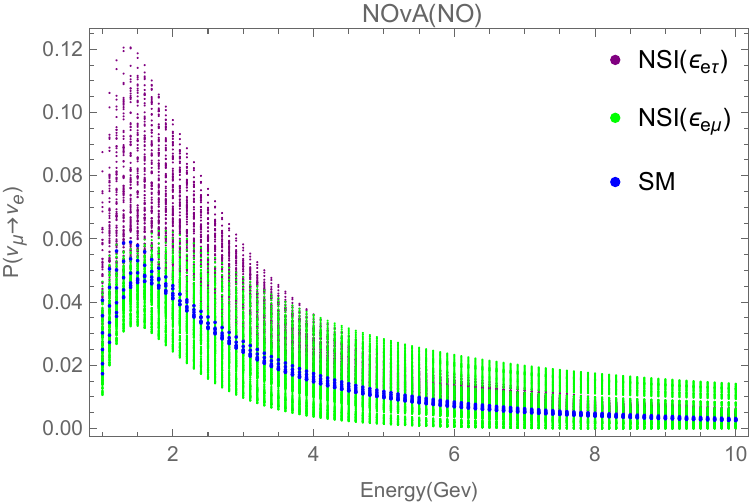}
\includegraphics[width=75mm]{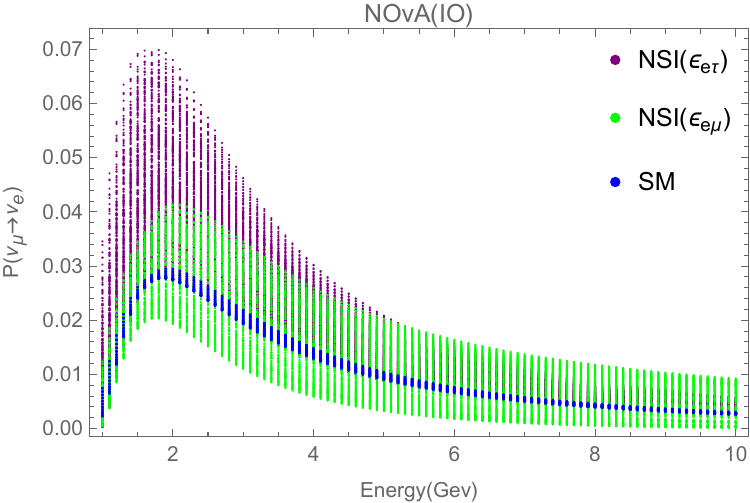}
\\
\includegraphics[width=75mm]{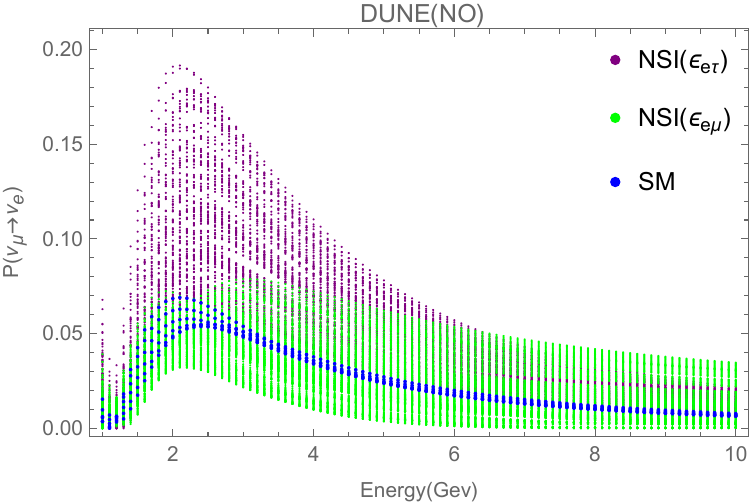}
\includegraphics[width=75mm]{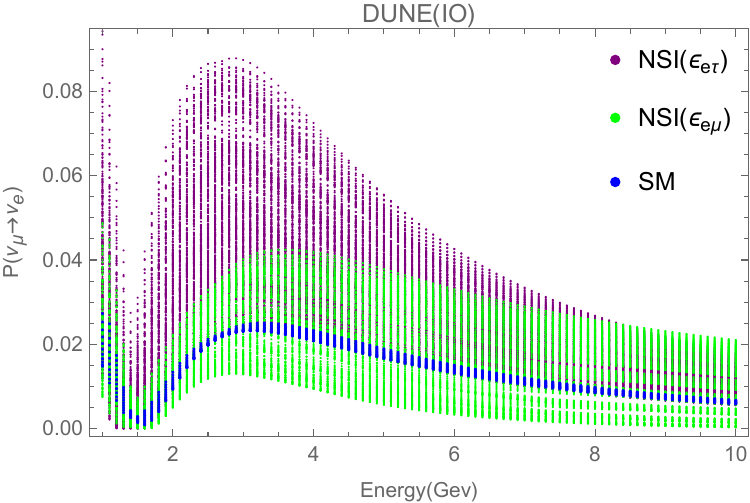}

\caption{ {\color{black} Probability vs. neutrino energy for NO (left) and IO (right) in the T2K (upper), NO$\nu$A (middle), and DUNE (lower) experiments. The green band represents LGtI with a $1\sigma$ allowed range of $\epsilon_{e\mu}$, $\phi_{e\mu}$, and cp phase, the purple band represents LGtI with $1\sigma$ allowed range of $\epsilon_{e\tau}$, $\phi_{e\tau}$ and cp phase, and the blue band represents LGtI with $1\sigma$ allowed range of cp phase. } }
\label{fig2}
\end{figure*}

{\color{black} Figure \ref{fig2} illustrates the variations in $P_{\mu e}$ when adjusting free parameters within $1\sigma$ range as a function of neutrino energy for SM and NSI scenarios in T2K, NO$\nu$A, and DUNE experiments. SM and NSI scenarios exhibit overlapping regions, so the effect becomes discernible primarily from the maximum deviation. For both ordering, the maximum deviation of $P_{\mu e}$ for the $\epsilon_{e\tau}$ scenario from SM is observed between $(1-3)$ GeV for NOvA and $(1.5-4.5)$ GeV for DUNE, while in IO, the $\epsilon_{e\mu}$ scenario also exhibits its maximum deviation within the same energy range. In the T2K experiment, it is hard to distinguish $P_{\mu e}$ for these scenarios for both NO and IO.

For NO, the maximum deviation for the $\epsilon_{e\tau}$ scenario from SM, in DUNE is $\approx$ 0.12 and $\approx$ 0.45 for $P_{\mu e}$ and $K_{3}$, respectively, while in NO$\nu$A, this deviation is $\approx$ 0.06  for $P_{\mu e}$ and $\approx$ 0.23  for $K_{3}$. For IO, the maximum deviation for the $\epsilon_{e\tau}$ scenario from SM, in case of DUNE is $\approx$ 0.06 and $\approx$ 0.19 for $P_{\mu e}$ and $K_{3}$, respectively, similarly in NO$\nu$A, deviation is $\approx$ 0.04  for $P_{\mu e}$ and $\approx$ 0.1  for $K_{3}$. The maximum deviation for the $\epsilon_{e\mu}$ scenario from SM, for DUNE is $\approx$ 0.02 and $\approx$ 0.07 for $P_{\mu e}$ and $K_{3}$, respectively, while in NO$\nu$A, maximum deviation is $\approx$ 0.01  for $P_{\mu e}$ and $\approx$ 0.04  for $K_{3}$.

Probability serves as a fundamental measure in experimental observations, particularly in the context of neutrino oscillation studies where it is commonly measured. LGtI can be expressed in terms of probabilities and exhibits complementary behavior to probabilities, providing an additional avenue to study the NSI effects. Hence, by initially detecting the NSI effects through probability measurements, we can reinforce their existence through LGtI observations. This dual approach strengthens our understanding of NSI phenomena and enhances our ability to characterize their impact on neutrino behavior. One advantage is that the possible deviation of NSI from SM is more pronounced in LGtI than in probabilities.

}

\section{Conclusion}
\label{sec:4}
In this work, we studied the impact of $\epsilon_{e\mu}$ and $\epsilon_{e\tau}$ NSI scenarios on temporal correlation, which is measured by LGtI in the context of three-flavor neutrino oscillation, and compared the results with SM predictions across three distinct accelerator experiment setups. These two NSI scenarios have the potential to resolve tension emerged in NO$\nu$A and T2K data. We look for the implication of new physics, whether the $\epsilon_{e \mu}$ or $\epsilon_{e \tau}$ NSI scenarios can be distinguished or not. We observed that in NO in the presence of NSI with $\epsilon_{e\tau}$, LGtI-violation significantly enhanced in comparison to SM. The degree of enhancement in $K_3$ in DUNE surpasses that of NOvA. However, in the case of IO, LGtI violation is enhanced for the $\epsilon_{e\tau}$ scenario over SM interaction, while LGtI violation for the $\epsilon_{e\mu}$ scenario is not affected significantly. Therefore, an interesting result of this work is that if LGtI is violated in the DUNE experiment for IO at energy above $8.5$ GeV, this would be possible only for the $\epsilon_{e\tau}$ scenario. The results of LGtI-violation for different NSI scenarios differ remarkably for DUNE. { \color{black} $\epsilon_{e\tau}$ scenario is favored in the context of normal ordering as well as inverted ordering. The possible deviation of NSI from SM is more pronounced in LGtI as compared to probabilities.}

\bibliographystyle{apsrev4-2}

\title{References}

\end{document}